\documentclass[a4paper,11pt]{article}
\pdfoutput=1 % if your are submitting a pdflatex (i.e. if you have images in pdf, png or jpg format)

\usepackage{jinstpub} % for details on the use of the package, please see the JINST-author-manual
\usepackage{amsmath}
\usepackage{graphicx}
\usepackage{bm}
\usepackage{siunitx}
\usepackage{comment}
\usepackage{placeins}
\usepackage{booktabs}

\title{\boldmath Embedded Trigger Control System for Wave--Plasma Interaction Experiments in Large Volume Plasma Device}

\author[a,1]{P.~K.~Srivastava,\note{Corresponding author.}}
\author[a,b]{Deepak~Rajpurohit,}
\author[a,b]{A.~K.~Sanyasi}
\author[a,b]{and~Ayan~Adhikari}

\affiliation[a]{Institute for Plasma Research, Bhat, Gandhinagar 382428, India}
\affiliation[b]{Homi Bhabha National Institute, Mumbai 400094, India}

\emailAdd{pkumar@ipr.res.in,draj91297@gmail.com}

\abstract{Magnetic-field control and synchronization are essential requirements in laboratory plasma experiments investigating wave propagation and plasma instabilities. In this work, an Embedded Trigger Control System (ETCS) has been designed and implemented for synchronized operation of a pulsed Helmholtz coil system and a broadband whistler-wave exciter in the Large Volume Plasma Device--Upgrade (LVPD-U). The experimental configuration employs an electrostatically shielded loop antenna driven by a thyratron-switched pulse-forming network to launch broadband whistler waves in the frequency range of approximately 150~kHz to 5~MHz. Simultaneously, a pulsed Helmholtz coil capable of generating magnetic fields up to $\pm150$~G is used to modify the local magnetic topology and create transient magnetic-null configurations. The developed ETCS utilizes the controlled evolution of the Helmholtz circuit $L/R$ time constant through the introduction of a lumped series inductor, enabling precise magnetic-field ramping and dynamic timing control. A microcontroller-based trigger architecture determines the desired firing instant from the calibrated current evolution of the Helmholtz circuit and synchronizes wave excitation with a prescribed magnetic-field amplitude. The system extends the effective circuit time constant by more than an order of magnitude, allowing accurate and repeatable wave launching during controlled magnetic-field evolution. The implemented approach provides a simple, reliable, and versatile synchronization platform for wave--plasma interaction studies requiring precise temporal coordination between pulsed magnetic-field generation and wave excitation.}


\begin{document}
\hypersetup{pdfauthor={Author}, pdftitle={Embedded Trigger Control System for Wave--Plasma Interaction Experiments in Large Volume Plasma Device}}
\maketitle
\flushbottom

\section{Introduction}
Magnetic nulls~\cite{49,50,51} play an essential role in the phenomenon of magnetic reconnection~\cite{11,43,44,45}, where magnetic field lines are broken and subsequently reconnected following topological rearrangement. During this process, the topology of magnetic field lines is changed due to which a substantial amount of magnetic energy from field lines is released and converts into kinetic energy and thermal energy of plasma particles embedded in them; this phenomenon is often seen in the solar corona~\cite{52,53,54}, magnetotail of the Earth's magnetosphere~\cite{55,56}, laboratory plasmas~\cite{58,59,60} and also in fusion devices~\cite{57}. So, in order to design controlled laboratory experiments for the study of magnetic reconnection, dynamically producing a magnetic null region is an essential requirement and can be achieved by rapid temporal variation of the magnetic field generated by a Helmholtz coil placed in a uniform background field. Recently, LVPD ~\cite{1,2}has been equipped with a Helmholtz coil with a diameter of $500 mm$ to create different types of magnetic field configuration, including a magnetic null at its center in the presence of the LVPD axial magnetic field. To study the spatial and temporal evolution of the magnetic null, it is crucial to track the formation of the magnetic null point along with the variation of Helmholtz magnetic field along with simultaneous excitation of the loop exciter to launch an electromagnetic wave in a plasma. The propagation of Whistler waves in plasma is generally linked with the background magnetic field, which is also responsible for the radial confinement of the plasma. The scenario of whistler wave interaction in a dynamic magnetic null region is created by suitable ramping of the Helmholtz magnetic field to cancel the axial magnetic field of LVPD. A fast Helmholtz trigger circuit is designed to produce a magnetic null by canceling a magnetic field of up to $\sim \pm 20G$. This system accepts the necessary experimental inputs to deliver a synchronized exciter trigger at a precisely defined instant during the null-field evolution, facilitating detailed wave–plasma interaction studies. 

The experimental system comprises several sets of controls, power supplies, and instruments and is described below. A Helmholtz coil, $500 mm$ in diameter, with $10$ turns in two sections, resistance of $110 m \Omega$, and inductance of  $ \sim 30 \mu H$. An IGBT-based power supply, rated $1 kA/120 V$, fed by a capacitor bank, providing a $10 ms$ pulse designed for inductive loads to drive the Helmholtz coil. An exciter power supply delivering up to $\sim100A $ at $5 kV $ to a single-turn electrostatically shielded loop antenna (matched to $50\Omega$ with an integral termination $50\Omega$, producing $100A$ pulses. The pulse width generated from lumped pulse forming network (PFN) is $8.5 \mu s$, with $250 ns$ rise time and $750 ns$ fall time. A master trigger from the plasma experiment control room, synchronized with all relevant subsystems. A microprocessor-based timing and control system for launching the exciter pulse. A pulse-forming network (PFN) switched by a hydrogen thyratron discharging into the loop antenna at $5 kV$, matched to $50 \Omega$ .

The Helmholtz coil current is ramped using an inductive time-constant elongation method to selectively trigger the exciter at specific energization levels. This is achieved by an inductor of $11.74mH$, and this generates an approximately linear field ramp of $\sim 10 G$ in $10 ms$.
An Arduino Uno microcontroller controls and triggers the Helmholtz and exciter power supplies. Its timers, counters, and registers generate delays from $10\mu s$ to $50 ms$ with high accuracy and repeatability. Desired data is set into the programmed micro-controller according to experimental requirements via a joystick and  $20\times 4$ LCD displays, eliminating the need for a computer. The joystick writes data to digital registers, enabling flexible sequential synchronization between the power supplies and the master plasma pulse trigger. Once configured and programmed, this standalone triggering system does not require a host computer or driver to adjust delay settings. This system performs satisfactorily in an electrical current switching environment of $\sim 1000A @ 10 \mu s$ that generates an electric field of $\sim 30V/m$ of induced interference within a typical unavoidable current circuit loop of $10cm^2$ area, which is an acceptable EMI scenario due to IGBT switching in the same electronic chassis.  

\section{Experimental Setup}

Figure 1 shows the schematic of the experimental setup of large volume plasma device. The large volume plasma device~\cite{1,2} (Fig. 1) has a length of $3m$ and a diameter of $2m$ offers a very conducive boundary free experimental region where whistler waves have been excited through loop antennas immersed within the plasma and studied under static background axial ($B_z$) magnetic field conditions. The device produces a moderately dense plasma having a plasma Density of $n_e \sim 10^{16} /m^3 $ to $  10^{18} / m^3 $, electron temperature: $T_e \sim 3-5 eV$, negative floating potential, $V_f <0$. The experimental zones are defined as i) Source region that contains the multi-filamentary plasma source, ii) EEF region that is produced by virtue of an diagnostics called electron energy filter (EEF) and the target region that received diffused plasma from the source region through EEF region. The axial thickness of the EEF region is $\sim 4cm$. The system is operated with an axial magnetic field, $B_z \sim 6.2G$ for producing the plasma. The system is also equipped with a Helmholtz coil having diameter of $50cm$, a $80mm$ diameter single loop antenna ($Z = 50\Omega$) and is excited with a $20kV$ pulse forming network with loop antenna excitation of $\sim 100A$ for a pulse duration of $\sim 8.5\mu s$. The pulse plasma is produced with the breakdown of the neutrals of the argon gas by impact ionization, and the duration of the plasma pulse is controllable between $5ms - 50ms$. 

\begin{figure}[tbp]
    \centering
    \includegraphics[width=0.95\textwidth]{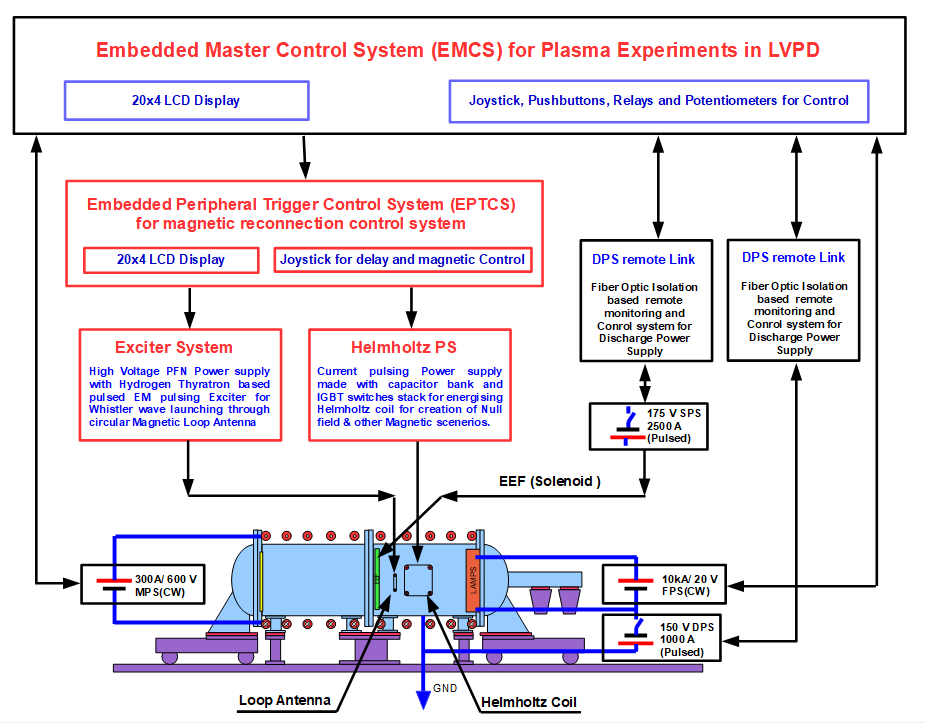}
    \caption{Large volume plasma device showing the Helmholtz coil, loop exciter antenna and associated power supplies, auxiliary power supplies used for plasma production and master trigger control for plasma experiments.}
    \label{fig:fig1}
\end{figure}

Figure 1 briefly presents various sub-systems, systems, power supplies and controls associated with the LVPD system. The system is equipped with $300A - 600V$ continuous mode operation magnet power supply (MPS) to charge the axial magnetic field coils, a $10kA - 20V$ continuous mode power supply to heat the tungsten filaments fixed to the plasma source (FPS), a $70V - 1kA$ pulsed mode ($\Delta t \sim 50ms$) power supply to produce plasma discharge (DPS). A thyratron based exciter circuit (Fig.2) is used for pulsing the loop antenna to generate electro-magnetic wave. A SCR based power supply for charging the Helmholtz coil (Fig.3), A master control triggers the filament, discharge power supplies along with a micro controller based magnetic field trigger circuit. Embedded Master Control System (EMCS), as mentioned in figure 1, is also based on Arduino-UNO micro-controller but it has been dedicated to the LVPD’s Plasma production involving various power supplies, e.g. discharge, filament, magnet and solenoid power supplies apart from generating trigger for DAQ and other secondary triggering systems like ETCS discussed here for Helmholtz and exciter power supplies triggering functions.  

\begin{figure}[tbp]
    \centering
    \includegraphics[width=0.85\textwidth]{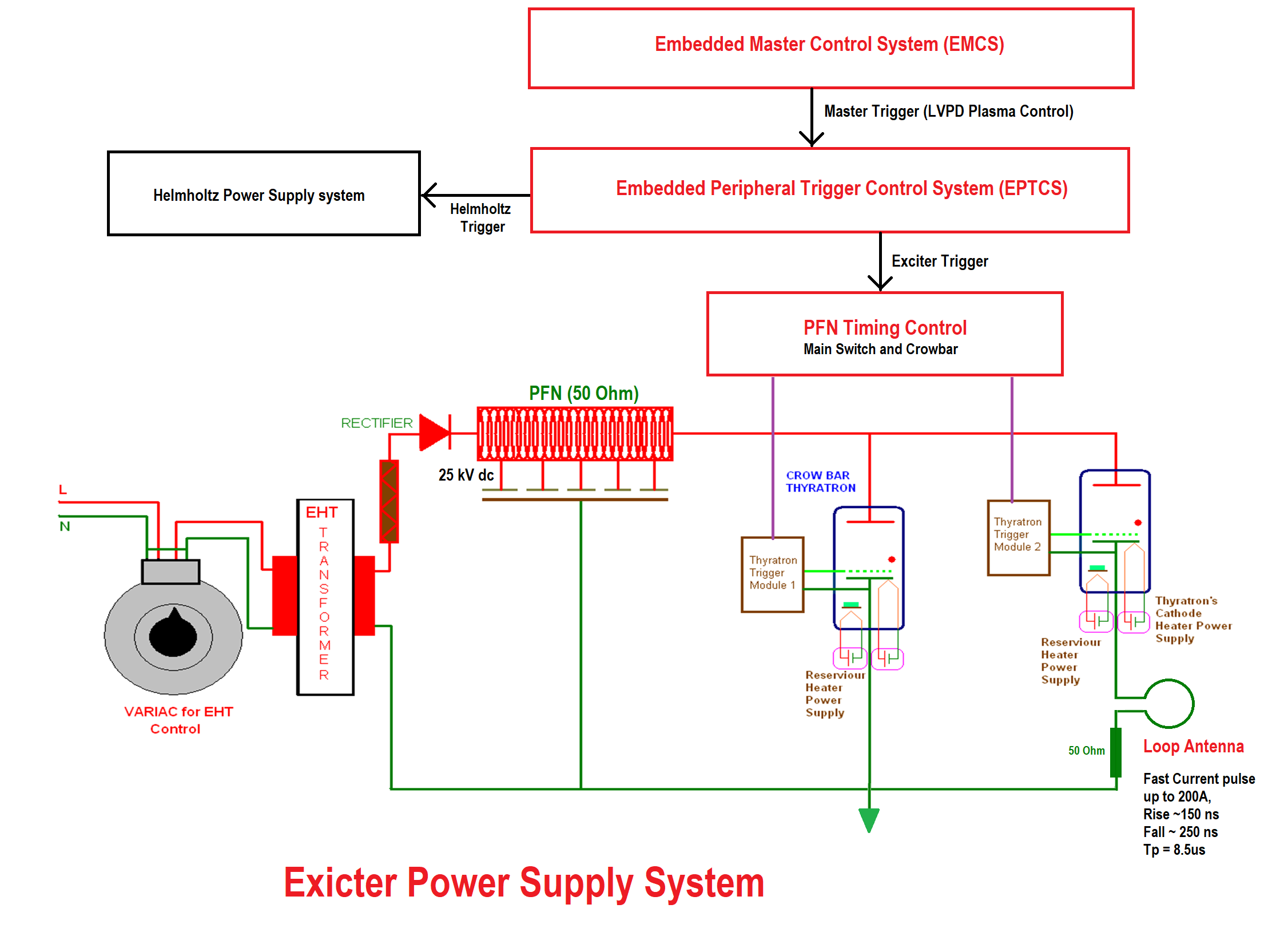}
    \caption{Exciter System.}
    \label{fig:fig2}
\end{figure}

\begin{figure}[tbp]
    \centering
    \includegraphics[width=0.95\textwidth]{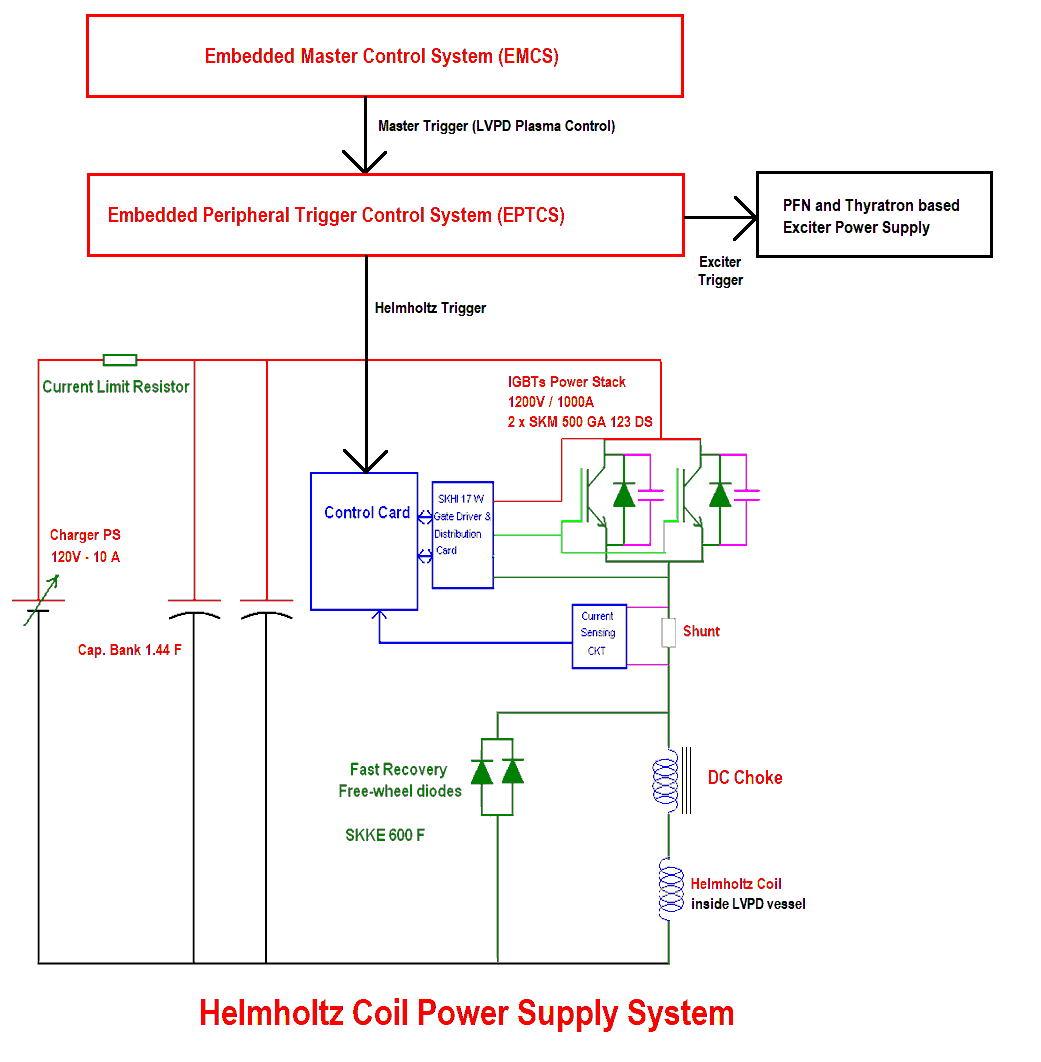}
    \caption{Helmholtz coil power supply system.}
    \label{fig:fig3}
\end{figure}

\subsection{Helmholtz coil field simulation and trigger}
\begin{figure}[tbp]
    \centering
    \includegraphics[width=0.95\textwidth]{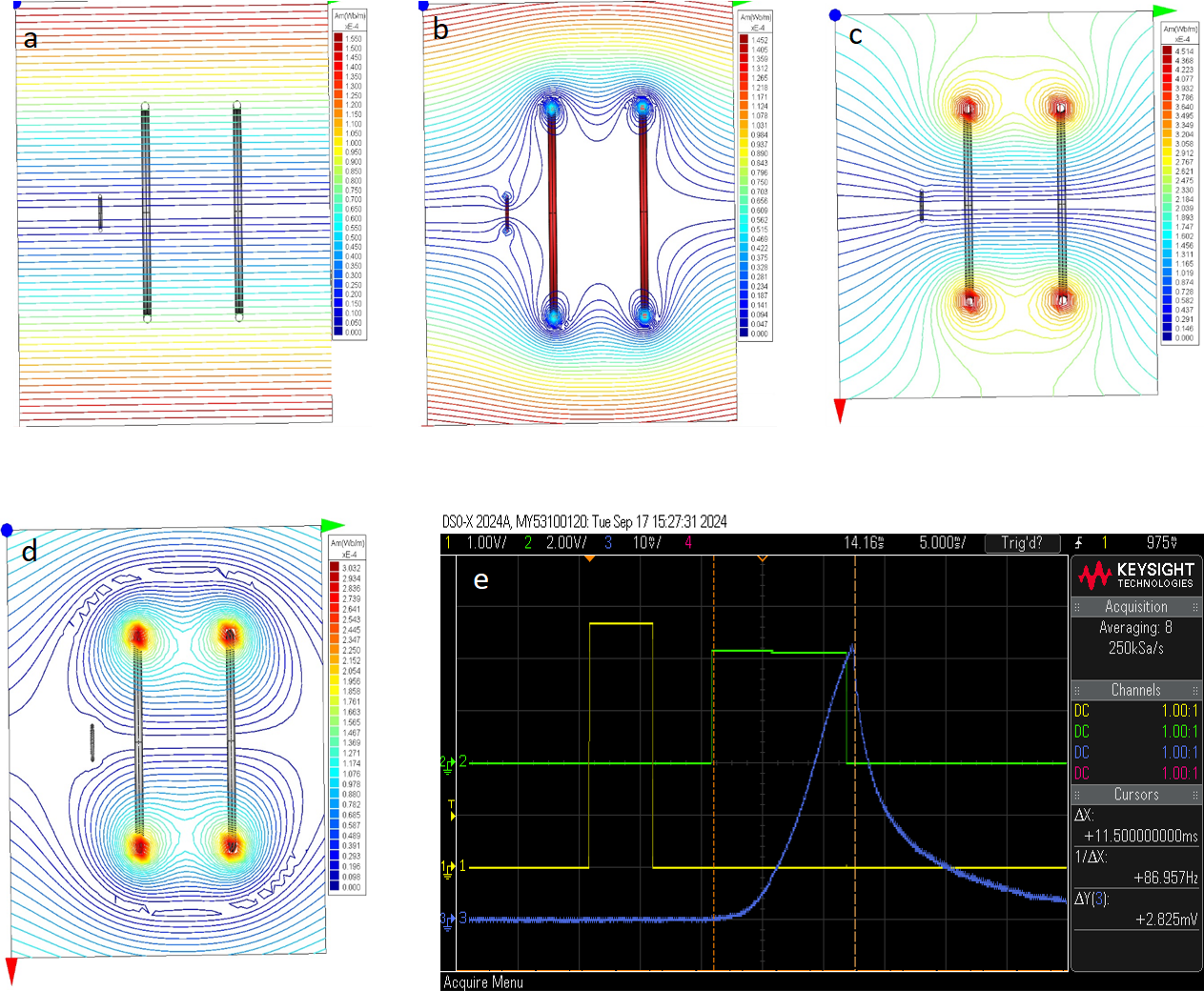}
    \caption{Simulated magnetic field patterns (Ampere code) of the Helmholtz coils along with the field generated due to the loop antenna current is shown, a) uniform axial magnetic field of LVPD without loop antenna, b) Helmholtz field produced a magnetic null region along with the loop antenna excitation, c) uniform magnetic field generation by the Helmholtz coil in the same direction along with loop excitation and the LVPD axial magnetic field, d) Helmholtz magnetic field produced opposite to the axial magnetic field of LVPD and e) master trigger pulse (yellow), delayed trigger (green) to Helmholtz activation and Helmholtz ramp-up current (blue).}
    \label{fig:fig4}
\end{figure}

The combined magnetic field of Helmholtz coil and loop exciter antenna installed in the LVPD system is simulated for its magnetic field profiles in different charging scenarios using the Ampere electromagnetic solver. Figure 4a shows the schematic of the Helmholtz coil installed in a simulated uniform axial magnetic field of LVPD; Figure 4b shows the magnetic null configuration within the Helmholtz volume by producing an exact magnetic field opposite to the axial magnetic field of LVPD and the loop antenna also excited. Figure 4c shows that Helmholtz field of $20G$ produced along the axial magnetic field of LVPD and figure 4d shows the $20G$ Helmholtz field produced opposite to the axial magnetic field of LVPD. Figure 4e shows the master trigger pulse to produce plasma (yellow), trigger (green) to the Helmholtz coil to activate it and the ramp-up current rise to the Helmholtz coil (blue) because of the $L/R$ time constant of the Helmholtz inductor coil. \\

This Helmholtz coil which works like a solenoid can be energized for axial magnetic field within itself as the same direction of $B_z$ or opposite to $B_z$ or to even cancel it to produce a null field at the center. The flexibility to produce a composite magnetic field scenario based upon the electrical current feed direction to the Helmholtz coil arms and magnitude to create various magnetic field line geometries for different transport conditions.
Slow sweep of the current amplitude in the Helmholtz coil $\delta t_{s} \sim 10 ms$ brings about a dynamic reconnection scenario in which the fast exciter $\delta t_{exc} < 8 \mu s$ can be turned on at desired time scales to generate composite magnetic field background information derived with measurements and also empirical estimations with magnetic field solvers.
A system capable of timing the exciter that produces a magnetic field produced within the Helmholtz coil by ramping up the current amplitude in controlled manner with an in line Inductor is designed and deployed in LVPD. The nonlinearity of the Inductive time constant of the ramping is normalized by measurements taken before in order to compensate the lookup table which has the information of relation of elapsed / delay time with Magnetic field ramped up in Gauss.

\subsection{Operation Principle}

In an experimental scenario described above, we have chosen micro-controller based electronic timing circuits with features like simple and interactive graphical user interface ( $20\times4$) LCD display, Joystick for setting up the required delay or magnetic field parameters with output on BNC connector based trigger that is taking synchronization command from Master Plasma Control to trigger the excitation of Helmholtz and Exciter system. The delay can be set to trigger the Helmholtz power supply pulsar circuit in microsecond time scales and since, it’s L/R time constant is regulated with the help of a series Inductor, the exciter antenna trigger is fired at a specific delay with the magnetic field gradually rises to a set value. Since, the non-linearity of this ramping slope is already known and which is recorded and fed into the micro-controller, the system is able to make a confident selection of setting parameter as it is already fed into the lookup table.  \\

\subsection{Methodology}
A precise triggering method was devised to investigate whistler-wave transmission and propagation along the background axial magnetic field ($B_z$) while simultaneously controlling the dynamic magnetic field generated by the Helmholtz coil ($B_H$). The Helmholtz coil is energized by an IGBT-switched capacitor bank ($1.44\text{ F}$, charged to $\sim 18\text{ V}$). In conventional direct-switched configurations, achieving fine, repeatable sub-Gauss magnetic-field increments across high-capacitance banks is challenging.

\begin{figure}[tbp]
    \centering
    \includegraphics[width=0.95\textwidth]{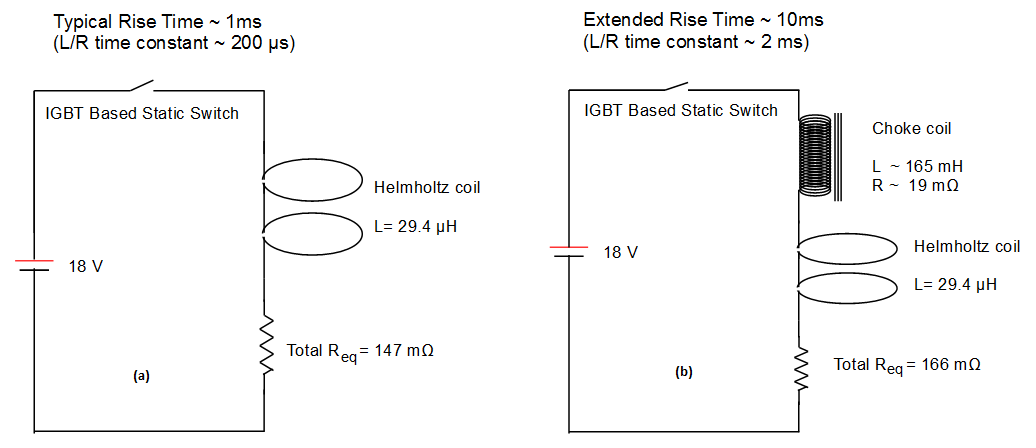}
    \caption{Schematic of the Helmholtz coil switching circuit for $L/R$ time-constant evaluation: (a) circuit configuration yielding an intrinsic time constant of $\sim 200\ \mu\text{s}$, and (b) modified circuit with an extended time constant realized via an auxiliary series DC choke inductor.}
    \label{fig:fig5}
\end{figure}

As shown in figure~\ref{fig:fig5}(a), the natural Helmholtz coil discharge circuit exhibits an inductance $L_0 \approx 29.4\ \mu\text{H}$ and an equivalent series resistance $R_{\text{eq}} \approx 147\text{ m}\Omega$, which yields an intrinsic electrical time constant:
\begin{equation}
\tau_0 = \frac{L_0}{R_{\text{eq}}} \approx 200\ \mu\text{s}.
\end{equation}
Under simple pulsed switching, the discharge current reaches its peak within approximately $1\text{ ms}$. This steep $dI/dt$ ramp makes selective, microsecond-scale triggering during the current rise impractical.

To address this, an inductive time-constant elongation method was adopted by inserting an auxiliary series choke inductor ($L_{\text{choke}} = 11.74\text{ mH}$, $R_{\text{choke}} \approx 19\text{ m}\Omega$) into the discharge path, as illustrated in figure~\ref{fig:fig5}(b). With the total circuit resistance becoming $R_{\text{total}} \approx 166\text{ m}\Omega$, the effective circuit time constant extends to $\tau = L_{\text{total}}/R_{\text{total}} \approx 71\text{ ms}$. Because the active IGBT conduction pulse duration is constrained to $\Delta t_{\text{pulse}} = 10\text{ ms}$ ($t \ll \tau$), the coil energization operates strictly within the initial, highly linear regime of the exponential charging profile:
\begin{equation}
I(t) = \frac{V_0}{R_{\text{total}}}\left(1 - e^{-t/\tau}\right) \approx \frac{V_0}{L_{\text{total}}}\, t \quad (\text{for } t \ll \tau).
\end{equation}
This establishes a repeatable, quasi-linear current ramp over $10\text{ ms}$ up to $\sim 20\text{ G}$. This predictable slope allows the operator to select any desired intermediate magnetic-field level and fire the transient high-voltage exciter during wave-propagation experiments, which are recorded within a $15\ \mu\text{s}$ acquisition window.

The series choke was implemented by utilizing the secondary winding ($6\text{ V}/25\text{ A}$) of an auxiliary low-voltage, high-current step-down transformer. This winding provides negligible ohmic resistance combined with a high magnetic saturation threshold. Core linearity is maintained up to $\sim 20\text{ G}$, which fully accommodates the field range required to dynamically nullify the LVPD axial background field ($B_0 = -6.2\text{ G}$). For higher field amplitudes ($> 20\text{ G}$), the series choke can be bypassed, allowing the circuit to operate in the fast-pulsing regime ($\sim 1\text{ ms}$ rise time) with magnetic amplitude set directly via the capacitor bank voltage.

\begin{figure}[tbp]
    \centering
    \includegraphics[width=0.75\textwidth]{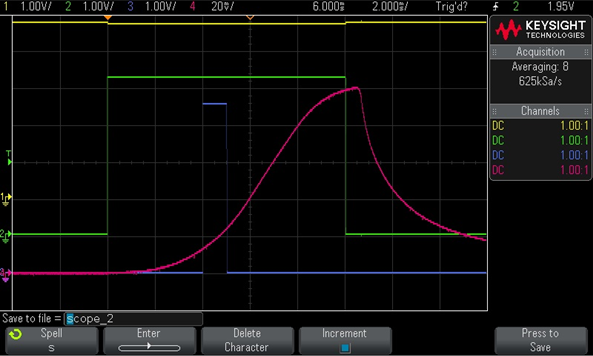}
    \caption{Oscilloscope screenshot showing the ETCS synchronization sequence: the external plasma trigger pulse (CH2, green) initiates the Helmholtz current ramp (CH4, magenta), followed by the delayed exciter firing trigger (CH3, blue) corresponding to the pre-selected magnetic-field amplitude.}
    \label{fig:fig6}
\end{figure}

The target magnetic field is selected by the user via the ETCS interface, and the microcontroller manages the synchronized trigger sequence. The required time delay between the initiation of the Helmholtz current and the exciter firing pulse is evaluated using a calibrated internal lookup table. Figure~\ref{fig:fig6} demonstrates the experimental timing sequence captured on an oscilloscope: the incoming plasma trigger (Channel 2, green) turns on the Helmholtz power supply. The ramped Helmholtz current (Channel 4, magenta) rises over $\sim 10\text{ ms}$ to produce a magnetic field of up to $20\text{ G}$ ($4\text{ G/div}$). After an elapsed delay of $\sim 4\text{ ms}$ (corresponding to a net intermediate field of $3\text{ G}$), the exciter trigger pulse is output on Channel 3 (blue trace) to launch the whistler wave packet.

\subsection{Programming Logic and algorithm flow chart of ETCS}

Figure 7 shows the flow Chart that displays sequential logical operations incorporated in the embedded controller based on Arduino-Uno micro-controller board. The program begins with initialization of standard input / output libraries for digital and analog signal pins along with I2C for LCD display panel. Initialization is done for various digital I/O and analog Input pins along with I2C bus as per system deployment plan shown in the table-1. 

After initialization of all the pins and functions, the system is prompted with a greeting message and asked to choose either Program mode or Run mode with the help of a toggle switch as per a brief message on the LCD panel.
If the Program Mode is selected, joystick lever is used with a facility to scroll and select two different timers, namely, Timer 1 and 2 by moving the cursor Up or Down. If timer 1 is selected, the delay time is in $\mu s$ from $1\mu s$ to about $15$ ms according to the experimental requirement.
Similarly, if Timer 2 is chosen by selecting joystick down, the time delay is set by horizontally sliding the same joystick lever. As the delay time is varied, the corresponding magnetic field value is also displayed at which the exciter is going to be fired. In this loop, the Micro-controller constantly displays the delay settings in response to the desired joystick operation which is programmed intuitively to manage desired delayed trigger settings.
Once the setting is over, the Mode selection toggle switch is shifted to Run Mode. In Run Mode, the hardware Interrupt gets activated and it stays in a small Quick refresh loop in which it just displays set values of timer 1, timer 2 and corresponding magnetic field value in Gauss till there is no plasma trigger generated (external). If the external Plasma trigger is received, the micro controller introduces an initial wait time in accordance to the time set in the timer 1 and then fires Helmholtz Pulsed power supply through a TTL pulse generated. \\

\begin{figure}[tbp]
    \centering
    \includegraphics[width=0.85\textwidth]{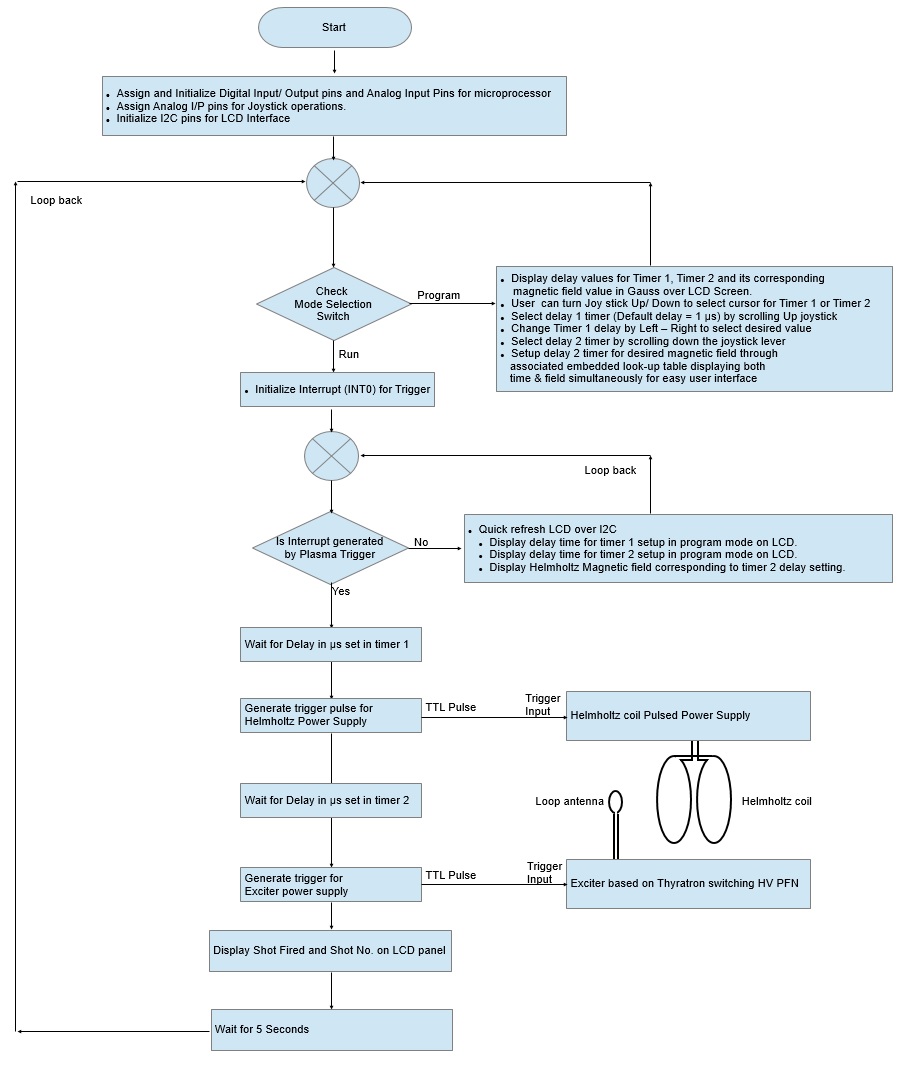}
    \caption{Flow Chart showing programming logic and algorithm of ETCS.}
    \label{fig:fig7}
\end{figure}

This process is further followed by another time delay which is introduced by the timer 2 setting during which the magnetic field of the Helmholtz keeps on ramping as per the standard L/R time constant progression already known and fed in the look up table. Once the delay of timer 2 is elapsed with subsequently reaching the desired magnetic field in the Helmholtz coil, another TTL pulse is generated which triggers the exciter system that launches a rectangular Pulse derived from a PFN on to a Loop antenna. This process of wave launching is as per the experimental needs where the wave is perfectly timed with a magnetic field for the reconnection studies.\\

This process occurs within few milliseconds, and is followed by a triggered shot message on the LCD display panel along Shot number fired and a wait for 5 Minutes before it can be re-triggered. The loop continues by again Re-Arming of Interrupt and waiting till next Master trigger pulse is received and the sequence is repeated. 

\begin{table}[tbp]
\centering
\caption{Microcontroller pin mapping and peripheral interface plan for the ETCS unit.}
\label{tab:pin_assignment}
\smallskip
\small
\begin{tabular*}{\textwidth}{@{\extracolsep{\fill}} c l c c p{6.8cm} @{}}
\hline
\textbf{Sr. No.} & \textbf{Signal Name} & \textbf{I/O Type} & \textbf{Pin No.} & \textbf{Functional Description} \\
\hline
1 & Mode Selection   & Input  & D2  & Toggle switch between PROG (+5\,V) and RUN (0\,V) modes \\
2 & Master Trigger   & Input  & D3  & External plasma discharge trigger command (Interrupt INT1) \\
3 & Helmholtz Trig.  & Output & D4  & 5\,V TTL trigger pulse for Helmholtz capacitor bank IGBT switch \\
4 & Exciter Trig.    & Output & D5  & 5\,V TTL trigger pulse for hydrogen thyratron PFN switch \\
5 & Mode LED         & Output & D6  & Status indication for RUN/PROG mode \\
6 & Trigger Status   & Output & D13 & Digital pulse indicator LED for triggered events \\
7 & Joystick X-axis  & Input  & A0  & Analog voltage input for horizontal cursor navigation \\
8 & Joystick Y-axis  & Input  & A1  & Analog voltage input for parameter/timer selection \\
9 & I$^2$C SDA       & I/O    & A4  & Serial data line for $20\times4$ character LCD module \\
10 & I$^2$C SCL      & Output & A5  & Serial clock line for $20\times4$ character LCD module \\
\hline
\end{tabular*}
\end{table}

\begin{figure}[tbp]
    \centering
    \includegraphics[width=0.6\textwidth]{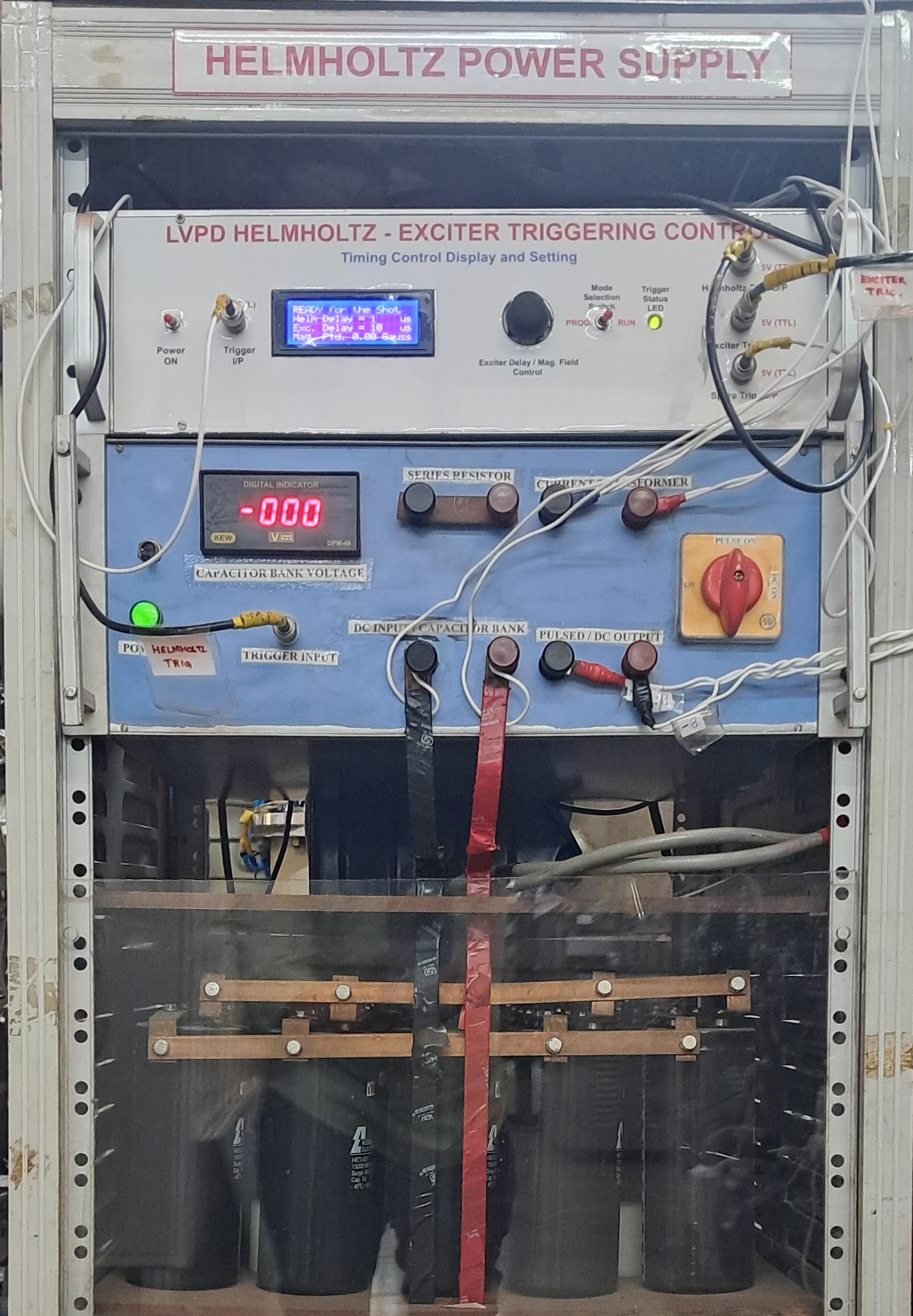}
    \caption{ETCS with Helmholtz power supply.}
    \label{fig:fig8}
\end{figure}

\begin{figure}[tbp]
    \centering
    \includegraphics[width=0.65\textwidth]{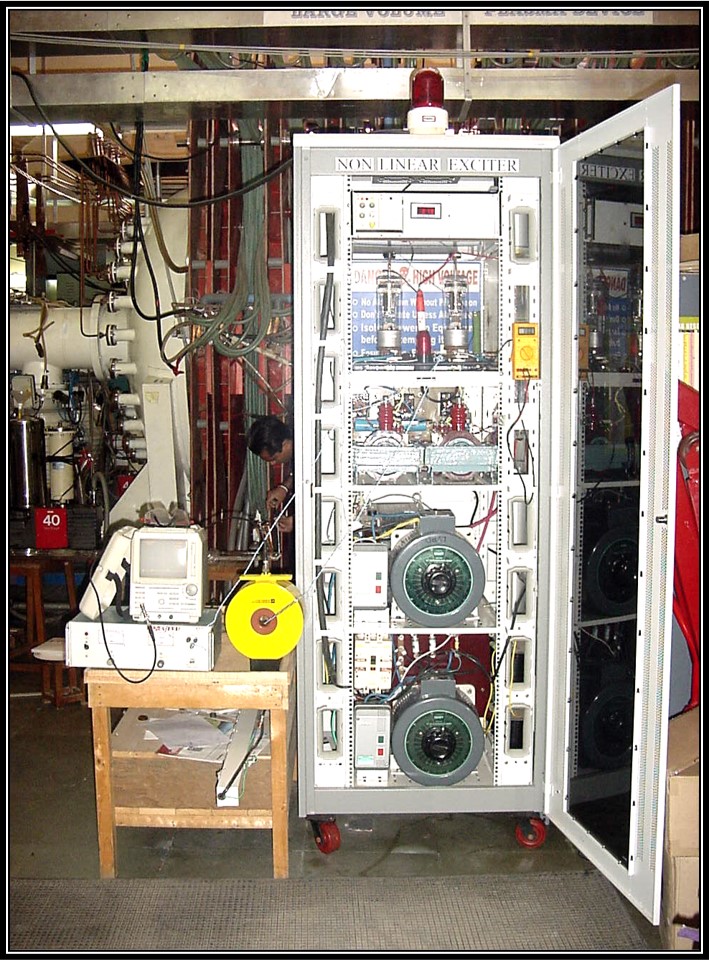}
    \caption{Exciter power supply having PFN and hydrogen thyratron switch.}
    \label{fig:fig9}
\end{figure}

\subsection{Controller Functionality}
Figure 8 shows the trigger power supply for the Helmholtz coil and loop exciter. Figure 9 shows the Thyratron based pulse forming network to pulsing the exciter loop. Figure 10 shows the face plate of the LVPD Helmholtz - Exciter triggering controller unit. The power ON switch controls the input power to the controller unit for initialization. The optically isolated triggering I/P is initiated from the experiment master plasma control setup with a $5V$ TTL pulse. The LED display is schematically represented as a blue rectangular patch where the system initial conditions are displayed. Modifications to the pulse control parameters are also displayed on this LED panel in real time. The mode selection switch operates in either PROGRAM or RUN mode. The PROGRAM mode allows manual variation in the timing control and delay settings for the exciter. The RUN mode facilitates the execution of the set parameters as I/P trigger is received. The exciter trigger delay and the current control to vary the Helmholtz magnetic field is done using a joystick with four degrees of movement and the push button fixes the parameter setting in the micro-controller. The trigger activation status is displayed on a dual colored (RED/GREEN) status LED light. The O/P triggers for the Helmholtz power supply and the exciter power supplies are generated at a predefined delay from the I/P trigger signal. The controller is equipped with a DC current transformer (CT) to measure the current passing through the Helmholtz coils which is a direct measure of magnetic field produced in Helmholtz coil. The logic of operation is explained schematically in figure 11.

\begin{figure}[tbp]
    \centering
    \includegraphics[width=0.95\textwidth]{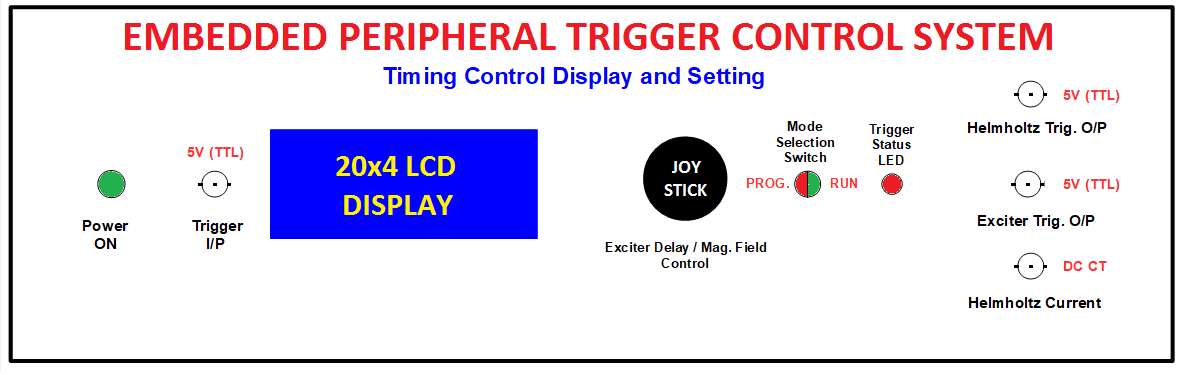}
    \caption{ETCS face plate showing trigger output for Helmholtz and exciter power supplies.}
    \label{fig:fig10}
\end{figure}
 
\begin{figure}[tbp]
    \centering
    \includegraphics[width=0.9\textwidth]{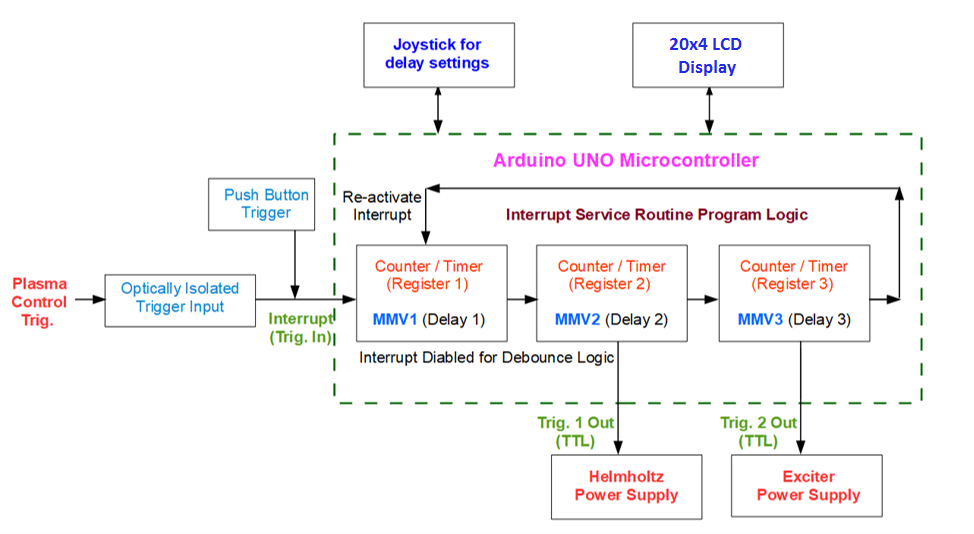}
    \caption{Schematic of ETCS programming logic tool.}
    \label{fig:fig11}
\end{figure}

\begin{figure}[tbp]
    \centering
    \includegraphics[width=0.65\textwidth]{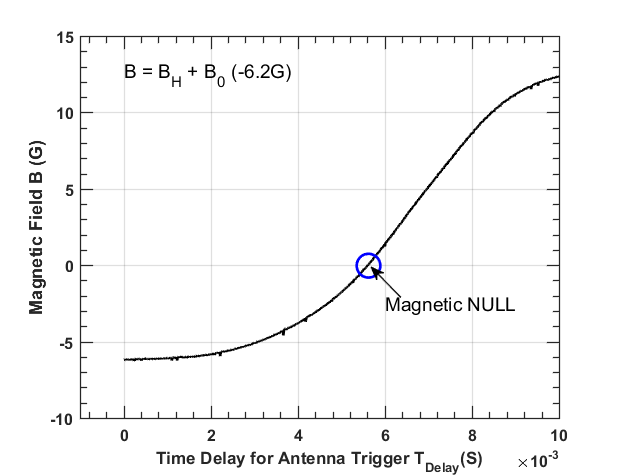}
    \caption{Magnetic field ramp-up using ETCS control with time.}
    \label{fig:EPTCS_measurement}
\end{figure}

Pulsed plasma experiments ($\Delta t\sim 50ms$) were carried out for wave plasma interaction studies in LVPD. The Helmholtz coil magnetic field ramp-up set from $0G$ to $18G$ within a time of $10msec$. The electromagnetic wave is launched from a loop antenna at a desired time delay with respect to the Helmholtz coil activation so that the measurements can be done at a pre-determined magnetic field at the center of the Helmholtz coil. Figure~\ref{fig:EPTCS_measurement} shows the data for the measured magnetic field produced at the Helmholtz coil center during the current ramp-up in the presence of LVPD axial magnetic field ($B_0 = -6.2G$). The inset circle corresponds to the time when the magnetic NULL is produced at the center of the Helmholtz coil. The data enables us to trigger the exciter to generate the electromagnetic pulse in the desired magnetic fields ($B = B_H + B_0$).\\

\section{Evolution of Magnetic Null with Ramping circuit}

\begin{figure}[h!]
  \centering
  \includegraphics[width=\textwidth,keepaspectratio]{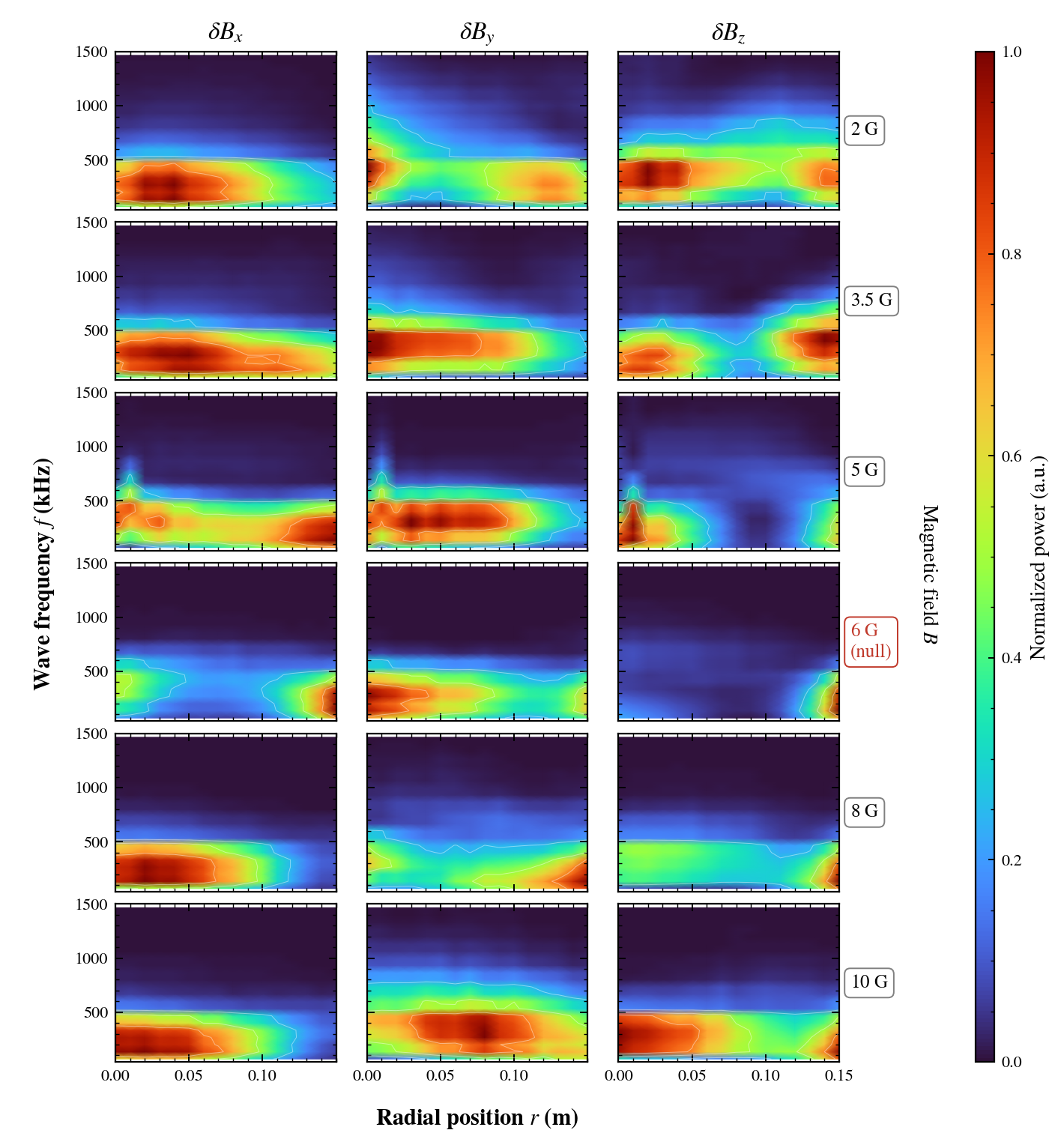}
  \caption{Radial--frequency maps of the normalized wave power for
  $\delta B_x$ (left), $\delta B_y$ (middle) and $\delta B_z$ (right) at
  background magnetic fields of 2, 3.5, 5, 6 (null), 8 and \SI{10}{G}
  (top to bottom). Each panel is normalized to its own minimum and maximum.}
  \label{fig:ramping}
\end{figure}
%\FloatBarrier

Figure~\ref{fig:ramping} presents the experimental wave power spectra measured at radial locations starting from the center of the Helmholtz coil ($r = 0.00\text{--}0.15\text{ m}$) with a spatial resolution of $\Delta r = 1\text{ cm}$. In this experiment, electromagnetic pulses are launched from a loop antenna into the LVPD plasma operating in an axial background magnetic field of $B_0 = +6.2\text{ G}$, upon which an opposing Helmholtz coil field ($B_H$) is imposed to control the net central magnetic field, $B = B_0 + B_H$. Static magnetic field configurations corresponding to $B_H = -2\text{ G}$, $-3.5\text{ G}$, $-5\text{ G}$, $\sim -6\text{ G}$ (establishing a magnetic null, $B \approx 0$, at the center), $-8\text{ G}$, and $-10\text{ G}$ are selected via trigger delay timing. 

The resulting electromagnetic fluctuations are recorded using a 3-axis bifilar magnetic probe ($\delta B_x$, $\delta B_y$, and $\delta B_z$) during loop antenna pulsation and processed via Fast Fourier Transform (FFT) to identify the dominant frequency bands. The radial profiles demonstrate a pronounced modification of wave propagation characteristics as the net magnetic field varies across the null point. Specifically, when $B \approx 0$ ($B_H \approx -6\text{ G}$), the $\delta B_x$ and $\delta B_z$ wave power components vanish entirely from the core region and shift radially outward, whereas the $\delta B_y$ component remains strongly localized near the center. While the detailed physical interpretation of these wave--plasma interactions is the subject of an ongoing study, these static-field measurements establish the foundation for current efforts to characterize magnetic fluctuations under dynamic ramping of the magnetic field from positive to negative polarity through the magnetic null.

%\FloatBarrier

\section{Discussion and Summary}

Embedded Trigger Control System (ETCS) has been conceptualized, instrumented, fabricated, configured and deployed in Large Volume Plasma Device experimental system for dynamically controlling the magnetic field of Helmholtz coil and loop exciter pulse for studying electromagnetic wave interaction with changing magnetic field scenarios. The control system is developed with the reasonably powerful Arduino Uno micro-controller boards and offers numerous user interfaces for making instruments in-house. Pulses of the order of $\sim 10 \mu s$ are easily achieved for triggering and setting up delays between functions of two different trigger circuits, i.e. the Helmholtz coil and the loop exciter. The magnetic field ramp up is set in such a way that the Helmholtz magnetic field tends to cancel the background uniform axial magnetic field of LVPD at the center of the coil thus produced a magnetic NULL and thereafter a dominating magnetic field opposite to the axial magnetic field. In the mean time, the loop antenna is triggered to generate an electromagnetic pulse that interacts with the plasma in a varying magnetic field. Presently, the ramp up time scale is fairly slow due to the L/R time stretching of the Helmholtz coil. Presently, the system is integrated to the LVPD experimental system and is being used in wave-plasma interaction studies in varying magnetic field scenarios including the magnetic NULL generation dynamically. 

\begin{comment}

\begin{figure}[tbp]
    \centering
    \includegraphics[width=0.6\textwidth]{Expt.PNG}
    \caption{Measured radial profile of wave power spectra for different magnetic field scenarios of Helmholtz configuration controlled by ETCS.}
    \label{fig:Expt}
\end{figure}
\end{comment}
% preamble: \usepackage{graphicx}  \usepackage{siunitx}

%%%%%%%%%%%%%%%%%%%%%%%%%%%%%%%%%%%%%%%%%%%%%%%%%%%%%%%%%%%%%%%%%%%%%%%%%%%%%%%%%%%%%%%%%%

\acknowledgments

The authors thank Dashrath Thakore for his significant contribution in the mechanical fabrication of the ETCS system along with active support of IPR workshop during fabrication of various cutouts of display panels, controls and input-output BNC terminations on the instrument chasis.  \\

\bibliographystyle{JHEP}
\bibliography{ref}

\end{document}